\documentclass[peerreviewca]{IEEEtran}
\IEEEoverridecommandlockouts
\usepackage{cite}
\usepackage{amsmath,amssymb,amsfonts}
\usepackage{algorithmic}
\usepackage{graphicx}
\usepackage{textcomp}
\usepackage{xcolor}
\def\BibTeX{{\rm B\kern-.05em{\sc i\kern-.025em b}\kern-.08em
    T\kern-.1667em\lower.7ex\hbox{E}\kern-.125emX}}
\begin{document}

\title{Community Detection Across Multiple Social Networks based on Overlapping Users\\
}


\author{\IEEEauthorblockN{Ziqing Zhu\IEEEauthorrefmark{1}, Tao Zhou\IEEEauthorrefmark{1}, Chenghao Jia\IEEEauthorrefmark{1}, Weijia Liu\IEEEauthorrefmark{2}, Jiuxin Cao\IEEEauthorrefmark{2}}
\IEEEauthorblockA{\IEEEauthorrefmark{1}\textit{School of Computer Science and Engineering},
\textit{Southeast University}, Nanjing, China\\
Email: zzqxztc@seu.edu.cn, zhoutao@seu.edu.cn}
\IEEEauthorblockA{\IEEEauthorrefmark{2}\textit{School of Cyber Science and Engineering},
\textit{Southeast University}, Nanjing, China\\
Email: jx.cao@seu.edu.cn}
}

\maketitle

\begin{abstract}
With the rapid development of Internet technology, online social networks (OSNs) have got fast development and become increasingly popular. Meanwhile, the research works across multiple social networks attract more and more attention from researchers, and community detection is an important one across OSNs for online security problems, such as the user behavior analysis and abnormal community discovery. In this paper, a community detection method is proposed across multiple social networks based on overlapping users. First, the concept of overlapping users is defined, then an algorithm CMN\_NMF is designed to discover the stub communities from overlapping users based on the social relevance. After that, we extend each stub community in different social networks by adding the users with strong similarity, and in the end different communities are excavated out across networks. Experimental results show the advantage on effectiveness of our method over other methods under real data sets.
\end{abstract}

\begin{IEEEkeywords}
community detection, online social networks, overlapping users, NMF
\end{IEEEkeywords}

\section{Introduction}
In recent years, with the rapid development of Internet technology and the large-scale popularization of smart mobile terminals, online social networks have gradually become important platforms for people's daily communication and information sharing.
However, the easy access and wide connection attributes of OSNs make it possible for those malicious users to achieve ulterior purpose, such as rumor spreading, spam comments, information theft and so on. It brings the severe security challenge to online social network services. At the same time, social network, as the online projection of human social relations in reality, contains huge data value. Many works take advantage of huge online social network data to discover effective method, which can analyse public opinions, monitor emergency, trace social interaction, etc. Among them, community detection work is an important research.

Social networks generally have cluster feature which is similar to the real society. This cluster feature can be defined by the concept of community. By analyzing the potential pattern behind the community, we can explore the information diffusion model, and discover the evolution and change rules in social network structure. The community reflects the correlation between individuals in the social network. The study of the community plays a crucial role in understanding the network structure, network function, information spread and other characteristics of networks.

Relevant studies\cite{xiang2016user} show that due to the differences in the functions of social networks, single social network is difficult to meet diverse demands of user, so users generally have more accounts on multiple social networks. Such users are called overlapping users across multiple social networks. The common existence of overlapping users enriches data information of single social network and plays an important role in information transmission across social networks, making it possible for many research works across OSNs such as behavior analysis and information diffusion, while the community detection on multiple social networks is the basis of that and plays an important role.

In this study, we propose a cross-social network community detection method based on overlapping users.  First, the overlapping user sets are extracted from multiple social networks, and then the overlapping user sets are divided into multiple stub communities in which users present close relevance by the CMN$\_$NMF, a community detection algorithm proposed in this paper. After that, each stub community stretches in different social networks by merging new users. Finally, the crossing social network communities are achieved once each stub community ceases growing.

The remainder of the paper is organized as follows. Section 2 reviews the related works. Section 3 gives the concept of overlapping users and the overlapping user network mode. The community detection method is analyzed in section 4. Section 5 evaluates the experiments. Finally, the last section concludes this paper¨s work and discusses the future research work.
\section{Related Work}
With the rapid development of big data technology and complex network research, community detection has become a popular topic in academic world.

In early research, many scholars pay attention to the study of community structure based on graph topology. Girvan et al.\cite{girvan2002community} proposed the method of community division using edge betweenness. The concept of modularity proposed by Newman et al.\cite{newman2004finding} becomes an important standard for evaluating community detection effect. Then Newman\cite{newman2004fast}, ClausetA\cite{clauset2005finding}, Blondel V D\cite{blondel2008fast} et al. proposed a variety of community detection algorithms by optimizing modularity. Considering that one member of a community can belong to different communities at the same time in the real world, Palla G et al.\cite{palla2005uncovering}\cite{derenyi2005clique} defined a complete subgraph structure with K vertices, called K-clique. Community structure is formed based on K-clique. Furthermore, community detection algorithm based on seed diffusion was proposed by Lancichinetti et al.\cite{lancichinetti2009detecting}, which taken the top N different users with the highest influence in the network as seed nodes, the community division is realized by calculating the closeness between ordinary nodes and seed nodes. Wang F et al.\cite{wang2011community} applied NMF (Nonnegative Matrix Factorization) to the work of community detection, and realized community partition of directed weighted network and undirected weighted network respectively.

In online social networks, there are many type relations between users, and only one-dimensional network relational data often fails to obtain the ideal community division. Many works focus on diverse data. Zhao et al.\cite{zhao2012topic} quantified the user's text content using TF-IDF in the Twitter network, together with  the user's follow relation to detect the community by K-mean clustering. Xu et al.\cite{xu2018community} built complex generation model based on the LDA (Latent Dirichlet Allocation) by integrating users' time, space, social and other relations, to solve and obtain the probability distribution of users in the location communities. Liu et al.\cite{liu2013multi} proposed a multi-NMF algorithm which can decompose dimensional data between nodes in the network to detect community. Through adding regularization term of multi-NMF algorithm, he et al.\cite{he2014comment} proposed two kinds of CoNMF community detection algorithm based on pair-wise and cluster-wise. Based on the Twitter network, Pei et al.\cite{pei2015nonnegative} constructed a three-layer relation matrix by integrating user-user relation, user-message relation and message-message relation, and used the joint non-negative matrix three-factor decomposition algorithm to obtain user membership in communities.

Furthermore, not just only focusing on single social network, a lot of research works are carried out for multiple social networks. Based on the idea of same users' similar clustering in different network, Nguyen H T. et al.\cite{nguyen2015community} fused different nodes in multiple networks into a unified model, using asymmetric nonnegative matrix decomposition algorithm to integrate network connection for all network, and obtained community. Philip S Y et al.\cite{philip2015mcd} studied the differences of community relations among overlapping users across social networks, and used spectral clustering to find community structure based on the confidence and discrepancy features of relations between users.

It is not difficult to find through the above introduction that the network community detection using matrix decomposition approach has become relatively mainstream technology. Meanwhile, related works do not fully explore the diversity of the association between overlapping users in multiple social networks, ignoring the limitations of matrix decomposition in the application of large social networks. In view of the above problems, this paper provides a more suitable method for cross-social network community detection.
\section{Conception and Model}
Some relative conception definitions and the hybrid network model will be first described here to better explain our research work; the details are as follows.

\emph{Definition 1}: \textbf{Multiple Social Networks}. This paper models multiple social networks as a collection $G$ of graphs. $G$ consists of $p$ single social networks. Each social network is expressed as  $G^i=(U^i,E^i)$ where $U^i$ and $E^i$ are the set of users and set of users' relations.

\emph{Definition 2}: \textbf{Overlapping Users}. One user can appear in different social networks. This paper defines users who have multiple accounts on different social networks as overlapping users.

\emph{Definition 3}: \textbf{Hybrid Network Model}. Considering overlapping users are crucial in community for information diffusion across social networks, this paper models a hybrid network based on overlapping users. The hybrid network is abbreviated by $G^o=(U^o,E^o)$, where $U^o$ is the set of all cross-social network overlapping users, and $E^0$ is the set of all overlapping users' association relations including ``follow'', ``comment'', ``like'' and ``semantic similarity'' relations. Based on the diversity of correlations, the overlapping user hybrid network $G^0$ is divided into four subnetworks $G^0=(G_1^0,\,G_2^0,\,G_3^0,\,G_4^{0i})$, where

$G_1^o=(U^o,\phi^1)$ is a subnetwork of follow relations, $\phi^1$ is the set of follow relation edge, each edge denotes at least one follow relation between two overlapping users in each social networks;

$G_2^o=(U^o,\phi^2)$ is a subnetwork of comment relations, $\phi^2$ is the set of comment relation edge, each edge is at least one comment relation between two overlapping users in each social networks;

$G_3^o=(U^o,\phi^3)$ is a subnetwork of like relations across social networks, $\phi^3$ is the set of like relation edge, each edge represents at least one like relation between two overlapping users in each social networks;

$G_4^oi=(U^o,\phi^{4i})$ is a subnetwork based on semantic similarity of post contents, $\phi^{4i}$ is the set of semantic similarity relation edge, each edge indicates semantic similarity of post contents between two overlapping users in $i^{th}$ social networks.

\emph{Definition 4}: \textbf{Stub Community}. This paper defines the communities which consist of overlapping users on hybrid network as stub communities, which are core basis of communities across multiple social networks.

Based on the above description, our research work of community detection across multiple social networks can be divided into the following two problems.

\emph{Problem 1}: Stub Community division based on hybrid network. Given hybrid network $G^o=(U^o,E^o)$, the task is to find the community membership for each overlapping user.

\emph{Problem 2}: Community extension in each single social network. Given each social network $G^i=(U^i,E^i)$, stub community $C^o$, the task is to add new users to this stub community, so a stub community will naturally grow up into a community across multiple social networks by stretching across different social networks.

Compared with other works for community detection across multiple social networks, our research has the following advantages:

1. A hybrid network model is specially designed for overlapping users to integrate the multiple dimensional information from different social networks, which benefits the community detection.

2. The stub community origins from overlapping user hybrid network, which is the same as the source of information diffusion across multiple social networks.

3. The community detection is divided into two phases, from the stub communities on overlapping users to communities across social networks, so the method is scalable for large social networks.
\section{Cross-social Network Community Detection}
\subsection{Stub community division algorithm}
Overlapping user nodes in hybrid networks present multidimensional association relations. In order to more effectively use these multidimensional relations, a \textbf{C}ross \textbf{M}ulti-\textbf{N}etwork community detection algorithm CMN\_NMF based on \textbf{NMF}(Nonnegative Matrix Factorization) is conceived deliberately.
NMF is a data dimensionality reduction method, which has been widely used in image analysis, text clustering and other fields, which has explainable, effective and flexible features. Here it's used to stub community detection.

NMF is a data dimensionality reduction method, which has been widely used in image analysis, text clustering and other fields, which has explainable, effective and flexible features. Here it's used to stub community detection.
\subsubsection{User adjacency matrix construction}
In the algorithm, four type weighted user adjacency matrixes are constructed, corresponding to the ``follow'', ``comment'', ``like'' and ``semantic similarity'' relations in the overlapping user hybrid network. Elements in the first three type adjacency matrixes are the number of follow, comment and like between users respectively across social networks. These adjacency matrixes represent asymmetric directed weighted networks. The elements in the last type matrix describe the semantic similarity between users of the topic distribution based on their post content, and this matrix is symmetric.
\subsubsection{Optimal Objective Function}
This paper proposes a stub community division algorithm CMN\_NMF for overlapping users by multi-dimensional hybrid network non-negative matrix decomposition.

In CMN\_NMF algorithm, the stub community membership matrix can be obtained by decomposing user adjacency matrixes. It can be seen from \eqref{objectionfunction} that the factorization for a asymmetric matrix is $U^{(t)}H^{(t)}(U^{(t)})^T$, for a symmetric matrix is $W^{(g)}(W^{(g)})^T$, here a common community membership matrix $S$ is used constraint condition to reach global optimization, and the matrix $S$ is the stub community division we want.
\begin{equation}
\begin{aligned}
\label{objectionfunction}
\min_{U^{(t)},W^{(g)},S}&{\sum_{t=1}^p a_t {\Arrowvert A^{(t)}-U^{(t)}H^{(t)}(U^{(t)})^T\Arrowvert}^2_F }\\
&+{\sum_{g=1}^q b_g {\Arrowvert X^{(g)}-W^{(g)}(W^{(g)})^T \Arrowvert}^2_F }\\
&+{\sum_{t=1}^p c_t {\Arrowvert(U^{(t)})^T U^{(t)}-S^T S\Arrowvert}^2_F }\\
&+{\sum_{g=1}^q d_g {\Arrowvert(W^{(g)})^T W^{(g)}-S^T S\Arrowvert}^2_F }\\
&s.t.  U_{ij}^{(t)} \geq 0,W_{ij}^{(g)})\geq 0,S_{ij} \geq 0,、\forall i,j
\end{aligned}
\end{equation}
where $a_t$ 、$b_g$ 、$c_t$ 、$d_g$ play the role of multidimensional association relation weight control. The elements in $S^T S$ represent cosine similarity between overlapping user community, which is understood as community similarity matrix. Compared with directly calculating the difference between community membership matrix  $U^{(t)}(W^{(g)})$  and result matrix $S$, calculating the difference between similarity matrix $(U^{(t)})^T U^{(t)}$($(W^{(g)})^T W^{(g)}$) and similarity matrix $S^T S$ can effectively reduce the influence of noise data. Other parameters are described in Tab.\ref{table_CMN_NMF_symbol_description}.

\begin{table*}[htbp]
\caption{CMN\_NMF symbol description}
\begin{center}
\begin{tabular}{ |c|c|c|}
\hline
\textbf{Symbol} & \textbf{Description}& \textbf{Dimension}\\
\hline
$p$&The number of directed network relations &-\\
\hline
$q$ &The number of undirected network relations &-\\
\hline
$k$ &The number of community in hybrid network &-\\
\hline
$n$ &The number of overlapping users &-\\
\hline
$A^{(t)}$& User adjacency matrix based on the $t^{th}$ directed relation &$n\times n$\\
\hline
$U^{(t)}$ & User-Community membership matrix based on the $t^{th}$ directed relation &$n\times k$\\
\hline
$H^{(t)}$ & Community-Community relation matrix based on the $t^{th}$ directed relation &$k\times k$\\
\hline
$X^{(g)}$ & Semantic similarity matrix of content topics among users in social network $g$ &$n\times n$\\
\hline
$W^{(g)}$ & User-Community membership matrix based on semantic similarity of text topics in social network $g$ &$n\times k$\\
\hline
$S$ &User-Community membership matrix based on fusion of all network relations &$n\times k$\\
\hline
\end{tabular}
\label{table_CMN_NMF_symbol_description}
\end{center}
\end{table*}
\subsubsection{Optimal Objective Function}
To solve the matrix $S$, it is required that the difference between the matrix $S$ and the user-community membership matrix under all relations is the minimum.

Aiming at the optimization objective function given in \eqref{objectionfunction}, the Lagrange multiplier method\cite{lee2001algorithms} is used to solves $U^{(t)},H^{(t)}, W^{(g)}$ and S respectively. At the same time, owing to the element value range is different between matrix $U^{(t)}$ and $W^{(g)}$, the non-zero diagonal matrices $Q^{(t)}$ and $ R^{(g)}$ are introduced to normalize each row element to the value range [0,1] to guarantee each row's element sum is 1. Here $Q^{(t) }=Diag(1/{\sum_{i=1}^k U_{1,i}^{(t)}},1/{\sum_{i=1}^k U_{2,i}^{(t)}},\ldots ,1/{\sum_{i=1}^k U_{n,i}^{(t)}})$, $R^{(g) }=Diag(1/{\sum_{i=1}^k W_{1,i}^{(g)}},1/{\sum_{i=1}^k W_{2,i}^{(g)}},\ldots ,1/{\sum_{i=1}^k W_{n,i}^{(g)}})$.

Based on \eqref{objectionfunction}, the Lagrange function is first given as follows:
\begin{equation}
\begin{aligned}
\label{lagrangeobjectionfunction}
\min_{U^{(t)},W^{(g)},S}&\sum_{t=1}^p \boldsymbol{\Big(}a_t {\Vert A^{(t)}-U^{(t)}H^{(t)}(U^{(t)})^T\Vert}^2_F \\
&+ c_t {\Vert(Q^{(t)}U^{(t)})^T (Q^{(t)}U^{(t)})-S^T S\Vert}^2_F \\
&-tr(\alpha ^{(t)}(U^{(t)})^T)-tr(\beta ^{(t)}(H^{(t)})^T)\boldsymbol{\Big)}\\
&+\sum_{g=1}^q \boldsymbol{\Big(}b_g {\Vert X^{(g)}-W^{(g)}(W^{(g)})^T \Vert}^2_F \\
&+d_g {\Vert(R^{(g)}W^{(g)})^T (R^{(g)}W^{(g)})-S^T S\Vert}^2_F \\
&-tr(\gamma^{(g)} (W^{(g)})^T)\boldsymbol{\Big)}-tr(\lambda S^T)\\
&s.t.  U_{ij}^{(t)} \geq 0,W_{ij}^{(g)}\geq 0,S_{ij} \geq 0,\\
&(\alpha^{(t)})_{ij} (U^{(t)})_{ij}=0,(\beta ^{(t)})_{ij} (H^{(t)})_{ij}=0,\\
&(\gamma ^{(g)})_{ij} (W^{(g)})_{ij}=0, \lambda _{ij} S_{ij}=0
\end{aligned}
\end{equation}

Then, let $ \frac{\partial J}{\partial U^{(t)}}=0$, $ \frac{\partial J}{\partial H^{(t)}}=0$ , $\frac{\partial J}{\partial W^{(g)}}=0$, $\frac{\partial J}{\partial S}=0$, and follow the KKT condition $(\alpha^{(t)})_{ij} (U^{(t)})_{ij}=0$, $(\beta ^{(t)})_{ij} (H^{(t)})_{ij}=0$, $(\gamma ^{(g)})_{ij} (W^{(g)})_{ij}=0$, $\lambda _{ij} S_{ij}=0$, the updating rules can be obtained, which are \eqref{update_U}, \eqref{update_H}, \eqref{update_W}, \eqref{update_S} respectively.

\begin{figure*}[!htb]
\begin{equation}
\label{update_U}
U^{(t)}\leftarrow U^{(t)}\odot \sqrt{\frac{a_t \Big(A^{(t)}U^{(t)}(H^{(t)})^T+(A^{(t)})^T U^{(t)}H^{(t)}\Big)+2c_t(Q^{(t)})^T Q^{(t)}U^{(t)}S^T S}{a_t\Big((U^{(t)}H^{(t)}(U^{(t)})^T)U^{(t)}(H^{(t)})^T+U^{(t) } (H^{(t) } )^T (U^{(t)} )^T U^{(t) } H^{(t)}\Big)+2c_t(Q^{(t)})^T Q^{(t)}U^{(t)}(Q^{(t)}U^{(t)})^T Q^{(t)}U^{(t)}}}
\end{equation}

\begin{equation}
\label{update_H}
H^{(t)}\leftarrow H^{(t)}\odot \sqrt{\frac{(U^{(t)})^T A^{(t)}U^{(t)}}{(U^{(t)})^TU^{(t)}H^{(t)}(U^{(t)})^TU^{(t)}}}
\end{equation}

\begin{equation}
\label{update_W}
W^{(g)}\leftarrow W^{(g)}\odot \sqrt{\frac{b_g X^{(g)}W^{(g)}+d_g(R^{(g)})^T R^{(g)}W^{(g)}S^T S}{b_g W^{(g)}(W^{(g)})^T W^{(g)}+d_g(R^{(g)})^T R^{(g)}W^{(g)}(R^{(g)}W^{(t)})^T R^{(g)}W^{(g)}}}
\end{equation}

\begin{equation}
\label{update_S}
S\leftarrow S \odot \sqrt{\frac{\sum^p_{t=1} c_t S(Q^{(t)}U^{(t)})^T Q^{(t)}U^{(t)}+\sum^q_{g=1} c_t S(Q^{(t)}U^{(t)})^T Q^{(t)}U^{(t)}}{\sum^p_{t=1} c_t S(S)^TS+\sum^q_{g=1} d_g S(S)^TS}}
\end{equation}
\hrulefill
\end{figure*}

Based on the iterative updating rulers, CMN\_NMF algorithm can be described below.
\renewcommand{\algorithmicrequire}{\textbf{Input:}}
\renewcommand{\algorithmicensure}{\textbf{Output:}}
\begin{algorithmic}[1] 
    \REQUIRE $p$ asymmetric user adjacency matrices, $q$ symmetric user adjacency matrices, Number of stub community $k$
    \ENSURE stub community membership matrix $S$
    \STATE Initialize the matrix $U^{(t)}$, $H^{(t)}$, $W^{(g)}$, $S$
    \WHILE {\eqref{objectionfunction} not converge}
    {
        \FOR{$t=1$ to $p$}
            \STATE Update $U^{(t)}$ according \eqref{update_U}
            \STATE Update $H^{(t)}$ according \eqref{update_H}
        \ENDFOR
        \FOR{$t=1$ to $p$}
            \STATE Update $W^{(g)}$ according \eqref{update_W}
            \STATE Update $S$ according \eqref{update_S}
        \ENDFOR
    }
    \ENDWHILE
\end{algorithmic}
\subsubsection{Convergence Analysis}
The iterative formula of parameters given in this study can guarantee the final convergence of the algorithm. The following is the proof of the convergence of $U^{(t)}$, this method is also applicable to other parameters.

\begin{equation}
\begin{aligned}
\text{Let} \quad L(U^{(t)})=a_t \Vert A^{(t) }-U^{(t)} H^{(t)} (U^{(t)})^T \Vert_F^2\\
+c_t \Vert(U^{(t) } )^T U^{t) }-S^T S\Vert_F^2
\end{aligned}
\end{equation}

Now $L(U^{(t)})$ is convex function with respect to $U^{(t)}$, and we need to prove that $L(U^{(t)})$ is nonincreasing function. That is:
\begin{equation}
L((U^{(t)})^{m+1} )\leq L((U^{(t)})^m )
\end{equation}
where $U^{(t)^x }$ represents the $U^{(t)}$ is obtained after the $x^{th}$ iteration with \eqref{update_U}.

Definition:

$G(U^{(t)},U^{(t)'})$is the auxiliary function of $L(U^{(t)})$, which meets the following conditions:
\begin{equation*}
G(U^{(t)},U^{(t)'} )\geq L(U^{(t)} )
\end{equation*}
\begin{equation*}
G(U^{(t)},U^{(t)} )=L(U^{(t)} )
\end{equation*}

Lemma:
\begin{equation*}
(U^{(t)})^{m+1}=\arg\min G(U^{(t) },(U^{(t)})^m )
\end{equation*}

Proof:
\begin{equation*}
\begin{aligned}
L((U^{(t)})^{m+1})\leq G((U^{(t)})^{m+1},(U^{(t)})^m )\\
\leq G((U^{(t)})^m,(U^{(t)})^m )=L((U^{(t)})^m)
\end{aligned}
\end{equation*}

First, the auxiliary function $G$ is given:
\begin{equation}
\begin{aligned}
&G(U^{(t)},U^{(t)'})=L(U^{(t)' })\\
&-2\boldsymbol{\Big(}a_t (A^{(t) } U^{(t)' } (H^{(t) } )^T+(A^{(t) } )^T U^{(t)' } H^{(t) } )\\
&+2c_t (Q^{(t)})^T Q^{(t) } U^{(t)' } S^T S\boldsymbol{\Big)} U^{(t)'} (\log U^{(t) } -\log U^{(t)' } )\\
&+2\boldsymbol{\Big(}a_t (U^{(t)' } H^{(t) } (U^{(t)' })^T U^{(t)' } (H^{(t) } )^T \\
&+U^{(t) '}(H^{(t)} )^T (U^{(t)' } )^T U^{(t)' }H^{(t) } \boldsymbol{\Big)}\\
&+2c_t \boldsymbol{\Big(}(Q^{(t) })^T Q^{(t) } U^{(t)' } (Q^{(t) }U^{(t)' } )^T Q^{(t) }U^{(t)'} )\\
&(\frac { (U^{(t) } )^2+(U^{(t)'} )^2}{2 U^{(t)'}}-U^{(t)' })\boldsymbol{\Big)}
\end{aligned}
\end{equation}

The Taylor expansion of the $L(U^{(t)})$ function is compared with the auxiliary function\cite{zhu2014tripartite} $G$.

When $U^{(t)'}=U^{(t)}$, $G(U^{(t)},U^{(t)'})=L(U^{(t)})$;

When $U^{(t)'}\neq U^{(t)}$, require $G(U^{(t)},U^{(t)'})\geq L(U^{(t)})$, only need to prove that:
\begin{equation*}
\begin{aligned}
&1.\quad -(U^{(t)}-U^{(t)'} )\leq -U^{(t)'}(\log U^{(t)}-\log U^{(t)'} )\\
&2.\quad U^{(t)} \leq \frac {((U^{(t) })^2+(U^{(t)'})^2)}{2(U^{(t)' } )}
\end{aligned}
\end{equation*}

The conclusions of these are easy to be proved ,and

$G(U^{(t)},U^{(t)'})\geq L(U^{(t)})$ is established, so that:
\begin{equation*}
\frac {\partial G(U^{(t)},U^{(t)'})}{\partial x}=0
\end{equation*}
Solving for $U^{(t)}$ and \eqref{update_U} can be obtained. Other iterative updating formulas are the similar as above.
\subsubsection{Reconstructing Relations in Single Social Network}
After obtaining all overlapping user stub communities divided in the overlapping user hybrid network, this study will use each overlapping user community as seeds to make community detection in different social networks, and dig out users who are similar to the overlapping user communities in a single social network.

The discovery strategy of community in single social network given in this study needs to calculate the connection strength between non-overlapping users and overlapping user communities, so a single social network is quantified as an undirected weighted network. In single social network, there are still a variety of association relations among users. However, the number of user nodes and relation edges in the network is large, in order to reduce network reconstruction time and improve the applicability of the algorithm, this study adopts the similarity of user follow relation to represent user similarity.

\begin{equation}
Sim_{i,j}=\frac {\text{intersection of user }i,j \text{'s friends set}}{\text{union of user }i,j\text{'s friends set}}= \frac {f_i \cap f_j}{f_i \cap f_j}
\end{equation}

The reconstruction method of single network relations are shown below.
\renewcommand{\algorithmicrequire}{\textbf{Input:}}
\renewcommand{\algorithmicensure}{\textbf{Output:}}
\begin{algorithmic}[1]
    \REQUIRE $G_{old}=(U,E_{follow} )$, $U$: represents all users of this network, $E_{follow}$: represent all edges of follow relation in this network
    \ENSURE $G_{new}=(U,E_{sim} )$, $U$: represents all users of this network, $E_{sim}$: represent all edges of user similarity in this network

    \STATE $E_{follsim}\leftarrow \phi$
    \FOR{each $u_i  \in U$}
        \FOR{each $u_j \in U$ }
            \STATE $e_{i,j}=\frac{f_i\cap f_j}{f_i\cup f_j}$
            \IF{$e_{i,j}>$ Threshold Value}
                \STATE $E_{follsim}.add (e_{i,j})$
            \ENDIF
        \ENDFOR
    \ENDFOR
    \STATE $E_{sim} \leftarrow \phi$
    \FOR{each $u_i  \in U$}
        \STATE $E_{i_neighbor}\in E_{follsim}$
        \STATE $L_i=ascending\_sort(E_{follsim})$
        \FOR{ $j$ to $\lceil \sqrt{\vert L_i \vert} \rceil$}
            \STATE $E_{sim}.add (e_{i,j})$
        \ENDFOR
    \ENDFOR
\end{algorithmic}

Through the reconstruction of network relational edge, each social network is reconstructed into undirected weighted network.
\subsection{Community Detection Algorithm based on Overlapping User Stub Community}
In this study, overlapping user stub community set $N^0=\{N_1^0,N_2^0,\dots,N_k^0\}$ and any reconstructed social network $G_{new}=(U,E_{sim})$ were obtained by reconstruct hybrid network stub community division and social network relations by overlapping users. Inspired by work\cite{pan2012detecting}, this paper used stub community $N_t^0 (1\leq t \leq k)$ to make community detection in the network $G_{new}$. It is necessary to calculate connection strength between user node $u_i$ outside subgroup $N_t^0$ and node $u_j$ in subgroup $N_t^0$ further for obtaining connection strength between user node $u_i$ and subgroup $N_t^0$.

The connection strength between user node $u_i$ and node $u_j$ in network $G_{new}$ was defined as $NS_{i,j}$, which was obtained by Dijkstra algorithm. The connection strength between any user node $u_i$ outside stub community $N_t^0$ and stub community $N_t^0$ is formulated as follows:

\begin{equation}
cl\_{NS_{i,N^0_t}}=\sum_{j \in N_t^0 }\frac {1}{size_{N^0_t}}NS_{i,j}
\end{equation}

The community threshold $Rec_t$ of stub community $N_t^0$ is defined. If $cl\_{NS_{i,N_t^0}}>Rec_t$, the user is put in community $C_{new}^t$ where $N_t^0$ is located.
The community detection algorithm based on overlapping user stub community are shown below.
\renewcommand{\algorithmicrequire}{\textbf{Input:}}
\renewcommand{\algorithmicensure}{\textbf{Output:}}
\begin{algorithmic}[1]
    \REQUIRE User set $U$ in network $G_{new}$, Connection strength set $cl\_{NS_{i,N_t^0}}$
    \ENSURE Community set $C_{new}$ in network $G_{new}$
    \FOR{each $N_i^0  \in N^0 $}
        \FOR{each $u_i \in (U-N_t^0)$ }
        \IF{$cl\_{NS_{i,N^0_t}}>Rec_t$}
            \STATE $C^t_{new}.add (u_i)$
        \ENDIF
        \ENDFOR
    \ENDFOR
\RETURN $C_{new}$
\end{algorithmic}

Through the above process, the community structure based on $k$ stub communities in each social network can be obtained. The communities with the same overlapping user in each social network are integrated across the network, and the user communities across the social network are finally obtained. The $k$ cross-social network communities is represented as $C^0=\{C_{1,}^0,C_2^0,\dots, C_k^0\}$.
\section{Experiments and Analysis}
\subsection{Data Description}
In this study, 1575 overlapping user accounts are crawled from Zhihu-Weibo as the data set. Zhihu is question-answer based social network, and Weibo is microblog based social network. The data statistics are as follows.

\begin{table}[!htb]
\caption{overlapping user data statistics}
\begin{center}
\begin{tabular}{ cc}
\hline
\textbf{Overlap user relation types} & \textbf{Number}\\
\hline
The follow relation&982\\
\hline
The comment relation&2065\\
\hline
The like relation&6243\\
\hline
\end{tabular}
\label{table_overlapping_user_data_statistics}
\end{center}
\end{table}

Based on the above overlapping user set, the additional non-overlapping user sets can be obtained by users' follow relation respectively in Zhihu and Weibo networks, the statistics on follow relations among users are shown as follows.

\begin{table}[!htb]
\caption{non-overlapping user data statistics}
\begin{center}
\begin{tabular}{ ccc }
\hline
\textbf{Social network} &\textbf{Number of users} &\textbf{Number of relations}\\
\hline
Zhihu&24294&2394472\\
\hline
Weibo&180736&5377427\\
\hline
\end{tabular}
\label{table_non_overlapping_user_data_statistics}
\end{center}
\end{table}

\subsection{Data Description}
\subsubsection{The number of community}
The number $K$ of community is a key parameter as an input of the algorithm CMN\_NMF, the reasonable setting of $K$ value plays an important role in the quality of community division. This study uses the Fast-Unfolding algorithm put forward by BlondelV\cite{blondel2008fast} to determine the $K$ value scope. Fast-Unfolding algorithm is based on given network by optimizing modularity to obtain the most appropriate number of community divisions.

FU algorithm only handle well topology based directed network. Here, we run FU algorithm 10 times in our experiment respectively on ``follow'',``like'' and ``comment'' subnetworks to get the optimal parameter $K$ by computing the average modularity. The details are shown in Table.\ref{table_average_number_of_community_division_and_modularity_based_on_FU_algorithm}.

\begin{table}[!htb]
\caption{Average number of community division and modularity based on FU algorithm}
\begin{center}
\begin{tabular}{ ccc}
\hline
\textbf{Subnetwork type} &\textbf{Number of communities} &\textbf{Modularity}\\
\hline
follow relation&15.6&0.38\\
\hline
like relation&15.1&0.27\\
\hline
comment relation&17.4&0.49\\
\hline
\end{tabular}
\label{table_average_number_of_community_division_and_modularity_based_on_FU_algorithm}
\end{center}
\end{table}

It can be seen from Table.\ref{table_average_number_of_community_division_and_modularity_based_on_FU_algorithm} that about 16 communities are best divided based on FU algorithm for each subnetwork in hybrid network. In order to ensure the experimental result reliability, the value range of $k$ is from 14 to 20.
\subsubsection{Evaluation Metrics}
There are two common evaluation metrics of community detection in complex network. The first is that the actual members of community are given in the network. The evaluation metric is to compare the difference between detected community and actual community. The second is to measure the community detection results by calculating the network modularity. In social networks, it is usually difficult to identify the members of community in the network, and the research objective in this paper is to find similar users across social networks. Based on the above considerations, the following two evaluation metrics are given in this paper.

\begin{itemize}
\item \textbf{Similarity of user text content}. Similar users have similar content interest. These similar users in different social networks are excavated out based on overlapping user stub communities, so the validity of algorithm can be verified by calculating the similarity of users' post content from different social network in the same community. Take Weibo and Zhihu data sets as examples. $C_1^0$ represents cross-social network community excavated from Weibo and Zhihu networks based on overlapping user stub community $N_1^0$. $C_1^w$ represents user set from Weibo social network in community $C_1^0$ (excluding overlapping users), and $C_1^z$ represents user set from Zhihu social network in community $C_1^0$ (excluding overlapping users). If the Weibo user set and Zhihu user set in same community have a high similarity, then the text content of the two user sets should have a high coincidence. This paper obtains text content form Weibo and zhihu networks respectively, calculate the word vectors in two user sets based on TF-IDF, and measure coincidence based on cosine similarity.
\item \textbf{Implicit overlapping user discovery}. An important basis for researches on community detection based on cross-social networks is that most overlapping users in cross-social networks have clustering similarity under different networks. For example, $u_i$ and $u_j$ are two overlapping users in Weibo and Zhihu, if $u_i$ and $u_j$ are in the same community in Weibo, then they also tend to be in  a  same community in Zhihu. Based on the similarity features of overlapping user clustering, this study provides a cross-network community evaluation metrics. Firstly, only two-thirds of seed overlapping users were randomly selected in each stub community to grow into cross-social network community. Then, it will find that the most remaining one-third overlapping users were hidden in the cross-social network as ``implicit overlapping user''. Because of the internal similarity in clustering, the ``implicit overlapping users'' are likely to appear in the same cross-network community. By calculating the frequency of such users in the cross-social network community, the quality of the cross-social network community can be indirectly reflected.

    This paper takes $C_t^0$, a cross-social network community based on t th overlapping user stub community $N_t^0$, as an example. The community is $C_t^w$ in Weibo and $C_t^z$  in Zhihu.

    Implicit overlapping user discovery ratio formula is as follows:
    \begin{equation}
        P_t=\frac{N_t^h}{\frac{1}{3} N_t^0}
    \end{equation}
    where $N_t^h$ represents the number of implicit overlapping user appearing in community $C_t^w$ and community $C_t^z$.
\end{itemize}
\subsubsection{Comparison Methods}
In order to verify the effectiveness of the cross-social network community detection method based on overlapping users proposed in this study, our method was compared with other methods. The methods involved in the experiment are introduced as follows:

\textbf{K-mean}: a fusion network was obtained by fusing the multi-dimensional relations of the overlapping user hybrid network, and the k-mean clustering algorithm was used to find stub community.

\textbf{ConcatNMF}: a fusion network was obtained by fusing the multi-dimensional relations in the overlapping user hybrid network, and the NMF clustering algorithm was used to find stub community.

\textbf{ColNMF}\cite{singh2008relational}: a common result matrix S is arranged, which is stub community membership matrix. NMF clustering algorithm was used to simultaneously find stub communities in each relation network by using one common matrix $S$. The formula is as follows:
\begin{equation}
\min \sum_{v=1}^{n_v} \lambda _v \Vert A^{(v)}-U^{(v)}S^T \Vert_F^2
\end{equation}

\textbf{MultiNMF}\cite{liu2013multi}: a public result matrix $S$ is arranged, which is stub community membership matrix. NMF clustering algorithm was used to find stub communities in each relation network respective, and public result matrix $S$ is used to associate each stub community matrix in every relation network. The formula is as follows:
\begin{equation}
\min \sum_{v=1}^{n_v}\Vert A^{(v) }-U^{(v)} (V^{(v)})^T \Vert_F^2 +\sum_{v=1}^{n_v}\lambda_v \Vert V^{(v) }-S\Vert_F^2
\end{equation}
\subsection{Results Analysis}
\begin{figure}[!htb]
\centerline{\includegraphics[width=0.52\textwidth]{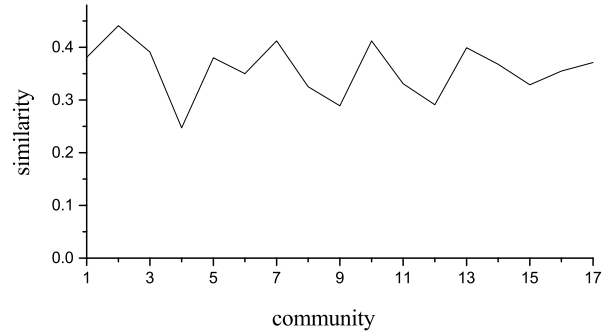}}
\caption{Similarity of text content between Weibo and Zhihu users in the community}
\label{fig_similarity_of_text_content_between_weibo_and_zhihu_users_in_the_community}
\end{figure}

When k value is 17, cross-social network community detection results based on overlapping users are obtained. 17 communities were divided, and the communities were sorted by the number of overlapping users from high to low. The similarity between Weibo users' text content and Zhihu users' text content in each cross-network community was calculated. The results are shown in Fig.\ref{fig_similarity_of_text_content_between_weibo_and_zhihu_users_in_the_community}.

\begin{figure}[!htb]
\centerline{\includegraphics[width=0.52\textwidth]{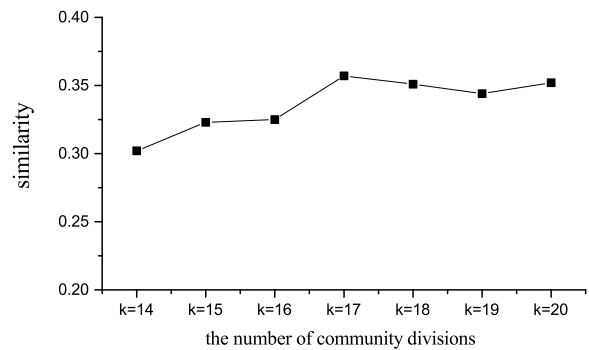}}
\caption{Average similarity of users text content with different k values}
\label{fig_average_similarity_of_users_text_content_with_different_k_values}
\end{figure}

When $k$ value is calculated as 14 to 20, the average text content similarity of WeiBo and Zhihu users in the community is calculated under each k value. The results are shown in Fig.\ref{fig_average_similarity_of_users_text_content_with_different_k_values}.

It can be seen from the figure that when the number of community divisions is greater than 16, the mean similarity of text content of Weibo and Zhihu users in all communities gradually tends to be stable.

\begin{figure}[!htb]
\centerline{\includegraphics[width=0.52\textwidth]{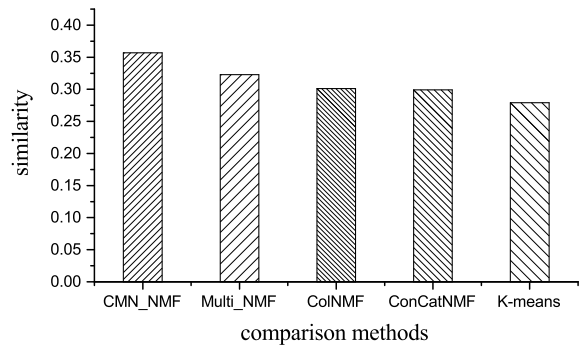}}
\caption{Comparison of average similarity of users text content with different methods}
\label{fig_comparison_of_average_similarity_of_users_text_content_with_different_methods}
\end{figure}

The average similarity of users' text content is shown in Fig.\ref{fig_comparison_of_average_similarity_of_users_text_content_with_different_methods}, which is calculated and compared with above described methods.

This figure shows the effectiveness of the stub community division algorithm proposed in the cross-social network community detection. Although MultiNMF have similar solution form based on NMF clustering algorithm compared with our method, it shows worse performance due to ignorance of text content specific to topic. The basic ideas of Co1NMF, ConcatNMF and k-mean algorithms are to directly fuse different types of user relations before stub community division. Data noise and dimensional difference and other problems have a significant impact on the results.

\begin{figure}[!htb]
\centerline{\includegraphics[width=0.52\textwidth]{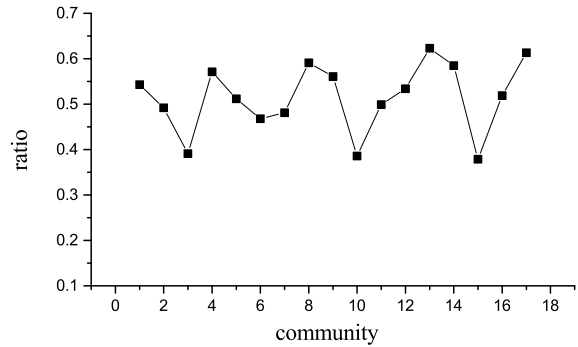}}
\caption{Discovery ratio of implicit overlapping user in single community(k=16)}
\label{fig_discovery_ratio_of_implicit_overlapping_user_in_single_community(k=16)}
\end{figure}

When k is set to 16, the stub community is found based on two-third of overlapping users for community detection in Weibo and Zhihu networks. The overlapping user ratio frequency of remaining one-third overlapping users in community is shown in fig.\ref{fig_discovery_ratio_of_implicit_overlapping_user_in_single_community(k=16)}.

\begin{figure}[!htb]
\centerline{\includegraphics[width=0.52\textwidth]{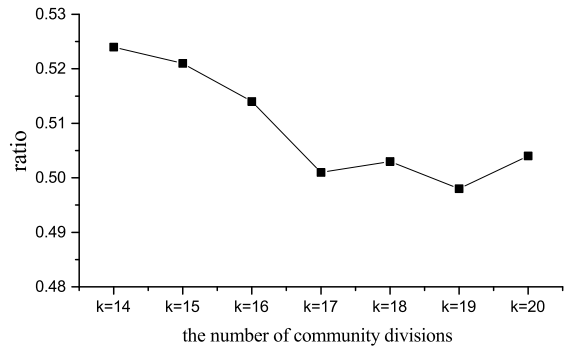}}
\caption{Average Discovery Ratio of implicit overlapping user under different k values}
\label{fig_average_discovery_ratio_of_implicit_overlapping_user_under_different_k_values}
\end{figure}

When k value is set 14 to 20, the average implicit overlapping user discovery ratio in community is shown in fig.\ref{fig_average_discovery_ratio_of_implicit_overlapping_user_under_different_k_values}.

The decrease of average rate reflects that the efficiency of implicit overlapping users' discovery will be affected by increase of the number of community divisions.

\begin{figure}[!htb]
\centerline{\includegraphics[width=0.52\textwidth]{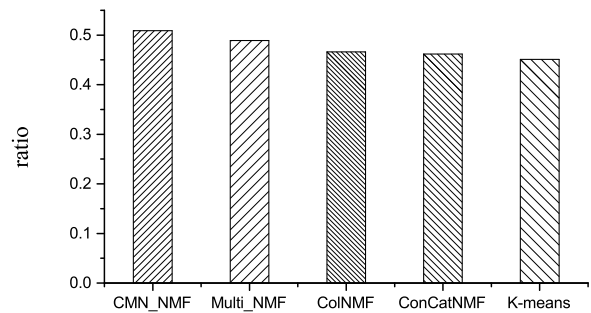}}
\caption{Comparison of average discovery ratio of implicit overlapping user with different methods}
\label{fig_comparison_of_average_discovery_ratio_of_implicit_overlapping_user_with_different_methods}
\end{figure}

The average discovery ratio implicit overlapping user is calculated, and compared with other algorithms. The results are shown in Fig.\ref{fig_comparison_of_average_discovery_ratio_of_implicit_overlapping_user_with_different_methods}.

The results demonstrate the effectiveness of hybrid network stub community division algorithm in community detection, and our method are more reliable than other algorithms to find overlapping users.
\section*{Acknowledgment}

This work is supported by National Natural Science Foundation of China under Grants No. 61772133, No.61972087.National Social Science Foundation of China under Grants No. 19@ZH014. Jiangsu Provincial Key Project  under Grants No.BE2018706.Natural Science Foundation of Jiangsu province under Grants No.SBK2019022870. Jiangsu Provincial Key Laboratory of Computer Networking Technology. Jiangsu Provincial Key Laboratory of Network and Information Security under Grants No. BM2003201, and Key Laboratory of Computer Network and Information Integration of Ministry of Education of China under Grants No. 93K-9.

\bibliographystyle{IEEEtran}
\bibliography{IEEEabrv,mybibfile}

\end{document}